# Optimization models of natural communication

Ramon Ferrer-i-Cancho[1]

ABSTRACT

A family of information theoretic models of communication was introduced more than a decade ago to explain the origins of Zipf's law for word frequencies. The family is a based on a combination of two information theoretic principles: maximization of mutual information between forms and meanings and minimization of form entropy. The family also sheds light on the origins of three other patterns: the principle of contrast, a related vocabulary learning bias and the meaning-frequency law. Here two important components of the family, namely the information theoretic principles and the energy function that combines them linearly, are reviewed from the perspective of psycholinguistics, language learning, information theory and synergetic linguistics. The minimization of this linear function is linked to the problem of compression of standard information theory and might be tuned by self-organization.

[1] Complexity and Quantitative Linguistics Lab. LARCA Research Group. Departament de Ciències de la Computació, Universitat Politècnica de Catalunya (UPC). Campus Nord, Edifici Omega, Jordi Girona Salgado 1-3. 08034 Barcelona, Catalonia (Spain). Phone: +34 934134028. E-mail: rferrericancho@cs.upc.edu.

1. INTRODUCTION

A family of information theoretic models of communication was introduced more than a decade ago to explain the origins of Zipf's law for word frequencies (Ferrer-i-Cancho & Solé, 2003; Ferrer-i-Cancho 2005a; Ferrer-i-Cancho & Díaz-Guilera 2007), understand the variation of the exponent of Zipf's law (Ferrer-i-Cancho, 2005b), explore the fatal consequences of an improper regulation of language principles and speculate on the origins of language (Ferrer-i-Cancho, 2006). Those models of Zipf's law suggest that "communication systems tend to be poised near critical states between phases of low memory effort and low disambiguation effort" (Kello & Belz, 2009).

Recent research has shown that models of that family are able to shed light on at least four apparently universal tendencies in world languages (Ferrer-i-Cancho, 2017; Ferrer-i-Cancho & Vitevitch, 2017; Ferrer-i-Cancho, 2016a). First, the principle of contrast, namely that *"every two forms contrast in meaning"* (Clark, 1987). Second, a vocabulary learning bias that is intimately related to the principle of contrast: the tendency of children to attach new words to unlinked meanings (Clark, 1987). Third, Zipf's meaning-frequency law (Zipf, 1945), that describes the dependency between $\mu$, the number of meaning of a word, and $f$, the frequency of a word, which has been approximated with (Ilgen & Karaoglan, 2007; Baayen & Moscoso del Prado Martín 2005; Zipf 1945)

$$\mu \propto f^{\delta}, \qquad (1)$$

where $\delta$ is a constant such that $\delta \approx \tfrac{1}{2}$. Fourth, Zipf's law for word frequencies, which states that the frequency of the *i*-th word of a text follows *approximately* (Zipf, 1949)

$$f \propto i^{-\alpha}, \qquad (2)$$

where $\alpha$ is a constant such that $\alpha \approx 1$. Empirical research on Zipf's law is initiating a new epoch by the intensive use of maximum likelihood for parameter fitting (e.g., Clauset, Shalizi, & Newman, 2009, Font-Clos, Boleda, & Corral, 2013), information theoretic model selection (Li, Miramontes & Cocho, 2010, Baixeries, Elvevåg, & Ferrer-i-Cancho, 2013, Gerlach & Altmann 2013) and a reorientation of the focus from the popular rank spectrum to the frequency spectrum (Clauset et al., 2009; Ferrer-i-Cancho & Gavaldà, 2009; Font-Clos et al., 2013). Empirical research on Zipf's law is also benefiting from the use of parallel corpora to control for variables such as genre or topic (Bentz, Kiela, Hill & Buttery, 2014) and the exploitation of very large corpora (Petersen, Tenenbaum, Havlin, Stanley, & Perc, 2011; Gerlach & Altmann 2013, Font-Clos et al., 2013). These trends are leading to a deeper understanding of the functional form of the law, for which Eq. 2 is just a rough approximation (Li et al., 2010; Font-Clos et al.; 2011, Gerlach & Altmann, 2013; Petersen et al., 2011; Bentz et al., 2014). A double law (a double power-law with two exponents) is becoming a stronger candidate for the organization of large corpora across communication systems in nature (Hernández-Fernández & Ferrer-i-Cancho, 2015; Gerlach & Altmann, 2013; Petersen et al., 2011; Ninio, 2005). This new research is reinforcing the view of Zipf's law as a robust pattern of language from different perspectives: the invariance of the law as text length increases, indicating a stable mould for word frequencies (Font-Clos et al., 2013), or the dependence of the law with respect to the linguistic units that are considered (Corral, Boleda, & Ferrer-i-Cancho, 2015). Finally, notice that Zipf's law can be seen both as an approximate pattern (Eq. 2) that needs explanation but also as a tool to capture differences between languages (Ha, Stewart, Hanna & Smith, 2006; Popescu & Altmann, 2008; Bentz, Verkerk, Kiela, Hill & Buttery, 2014).

In this article, a family of models (Ferrer-i-Cancho & Solé, 2003; Ferrer-i-Cancho, 2005a; Ferrer-i-Cancho, 2014; Ferrer-i-Cancho, 2017; Ferrer-i-Cancho & Vitevitch, 2017) that is able to shed light on these four patterns of language will be reviewed. It is worth mentioning that bidirectional optimality theory provides complementary accounts of diachronically stable and cognitively optimal form-meaning pairs that goes beyond the scope of the present article (Benz & Mattausch, 2011). Section 2 presents the family of models and details the aspects of the family to be reviewed. Sections 3-5 review the core components of the family from the perspective of psycholinguistics, language learning, information theory and synergetic linguistics. Section 6 reviews the need for self-organization. The article ends in Section 7 with a discussion on the construction of a general but parsimonious theory of natural communication.

2. A FUNCTIONAL INFORMATION THEORETIC FRAMEWORK

The mapping of forms into meanings of a communication system (e.g., human language) can be modelled by means of a matrix defining the weight of the association for every form-meaning pair (Steels, 2000; Ferrer-i-Cancho, 2005a). The joint probability between a form (e.g., a word) and meaning is a way of defining that weight. If $s_i$ is the $i$-th form of a finite repertoire of size $V_S^{max}$, and $r_i$ is the $i$-th meaning of a finite repertoire of meanings of size $V_R^{max}$, $p(s_i, r_j)$ is the joint probability of $s_i$ and $r_j$. For the particular case of human language, the matrix defined by $p(s_i, r_j)$ for every possible $s_i$ and $r_j$ can be seen as a two-way associative memory storing form-meaning pairs. The marginal probabilities are obtained from the joint probabilities as usual, i.e.

$$p(s_i) = \sum_{j=1}^{V_R^{\max}} p(s_i, r_j) \tag{3}$$

and

$$p(r_i) = \sum_{j=1}^{V_S^{\max}} p(s_j, r_i). \tag{4}$$

Let us define $\mu_i$ as the number of meanings of the $i$-th form, namely the number of meanings for which $p(s_i, r_j) > 0$. From the joint probabilities, the probability of forms (Eq. 3) and $\mu_i$, the four empirical laws reviewed in Section 1 can be investigated theoretically. If the principle of contrast holds perfectly and $p(s_i, r_j) > 0$, one expects that $p(s_k, r_j) = 0$ for $i \neq k$. The tendency of children to attach new words to unlinked meanings can also be stated in probabilistic language: if a new word arrives (a word such that $p(s_k=0)$) the child will tend to attach it to a meaning such that $p(r_l=0)$, converting $p(s_k, r_l) = 0$ into $p(s_k, r_l) > 0$. Zipf's meaning-frequency law (Zipf, 1945) can also be verified from $p(s_i)$ and the value of $\mu_i$ derived from the joint probabilities. One can check if Zipf's law holds for word probabilities once all the *p*($s_i$) are sorted decreasingly.

G. K. Zipf (1949) hypothesized that the law that bears his name (Eq. 2) revealed principles about the organization of vocabularies. Half a century later it is believed that the scaling laws arising at different levels and domains of cognition, with Zipf's law for word frequencies as an example among many (Kello et al., 2010), might be expressions of general principles (Kello, 2013; Ferrer-i-Cancho et al., 2013).

A decade ago, the translation of Zipf's arguments to information theory lead to two hypotheses about Zipf's law (Ferrer-i-Cancho & Solé, 2003; Ferrer-i-Cancho, 2005a):

- Strong hypothesis: Zipf's law could originate from the minimization of Ω, a cost function combining two principles, the maximization of *I*(*S*,*R*), the mutual information between forms and meanings, and the minimization of *H*(*S*), the entropy of forms. For the strong hypothesis, the minimization of Ω is a causal force.
- Weak hypothesis: Zipf's law could be optimal with regard to the minimization of Ω but it would not be a product of the minimization. This hypothesis means that the optimality of Zipf's law in terms of Ω would be achieved as a side-effect of other mechanisms.

Both hypotheses are abstract and do not involve principles that are necessarily specific to words or language. Therefore, they can be applied to other levels of language, e.g. inflectional paradigms, phonology (Kello & Beltz, 2009) and even to other species (McCowan, Doyle, & Hanser, 2002; Markov & Ostrovskaya, 1990) and the genetic code (Obst, Polani & Prokopenko, 2009). The generality of these models is neglected when reviewing models of Zipf's law for word frequencies (Piantadosi, 2014).

Interestingly, a challenge for the weak hypothesis is that the models of the family reviewed here make predictions beyond Zipf's law. This is a very important requirement of models of Zipf's law for word frequencies (Piantadosi, 2014). In particular, the family sheds independent light on the origins of the principle of contrast and the vocabulary learning bias above (Ferrer-i-Cancho, 2017; Ferrer-i-Cancho & Vitevitch, 2017) and a relaxed version of the meaning-frequency law, i.e. Eq. 1 with $\delta$=1 (Ferrer-i-Cancho, 2016). Those further predictions have not been considered when reviewing those models critically (Piantadosi, 2014) and provide support for the strong hypothesis.

The cost function Ω and its components will be defined in detail next. The entropy of forms is defined as

$$H(S) = -\sum_{i=1}^{V_S^{\max}} p(s_i) \log p(s_i). \tag{5}$$

A convenient definition of *I*(*S*,*R*) for theoretical purposes is

$$I(S,R) = \sum_{i=1}^{V_S^{max}} \sum_{j=1}^{V_R^{max}} p(s_i, r_j) \log \frac{p(s_i, r_j)}{p(s_i) p(r_j)}. \tag{6}$$

The particular form of the energy function that has been proposed is (Ferrer-i-Cancho, 2005a; Ferrer-i-Cancho & Solé, 2003)

$$\Omega(\lambda) = -\lambda I(S,R) + (1-\lambda) H(S), \tag{7}$$

where $\lambda$ is a parameter controlling for the weight of each of the two pressures, $0 \leq \lambda \leq 1$. If $\lambda=0$ then mutual information maximization is irrelevant and if $\lambda=1$ then form entropy minimization is irrelevant.

Although we are using human language as our guiding example, we believe that our framework does not need a mind or a brain. The theoretical framework presented is abstract enough to allow one to replace forms and meanings, respectively, by tags and resources (Cattuto, Loreto, & Pietronero, 2007), call types and behavioral contexts (McCowan et al., 2002; Ferrer-i-Cancho & McCowan, 2009), or codons and amino acids (Bel-Enguix & Jiménez-López 2011; Obst et al., 2009).

Our framework is a functional and information theoretic approach to language. Our claims about the efficiency of language are supported by two fundamental components:

- A clear statement of the energy function that languages hypothetically minimize (Eq. 7). Recently, the efficiency of language has been defended using quantitative and information theoretic approach but the energy or cost function that measures that efficiency is lacking (Piantadosi, Tily & Gibson, 2011). We do not mean that models of language require an energy function by default but that a cost or energy function (and its justification) is a requirement for claims on efficiency.
- Mathematical arguments (analytical, numerical or both) for the relationship between the energy function and the observed patterns of language. A necessary requirement for a well-defined theory of language efficiency is a derivation of patterns of language, e.g. Zipf's law for word frequencies or Zipf's law of abbreviation, from cost functions (Ferrer-i-Cancho et al., 2013; Ferrer-i-Cancho, 2005a, 2005c; Ferrer-i-Cancho & Solé, 2003). This is also missing in recent claims about the efficiency of language (Piantadosi, Tily & Gibson, 2011).

Here we address two important challenges for the hypothesis that languages and genomes (Ferrer-i-Cancho, 2017; Obst et al., 2009) minimize $\Omega(\lambda)$ directly (or indirectly):

- The rationale behind the two ingredients of $\Omega(\lambda)$: the minimization of $H(S)$ and the maximization of $I(S,R)$. Previous arguments will be improved and expanded with insights from psycholinguistics, information theory and synergetic linguistics.
- The rationale behind the way the two principles are combined. The definition of $\Omega(\lambda)$ looks arbitrary. The right way of combining $-I(S,R)$ and $H(S)$ does need to be linear. A theoretical justification of $\Omega(\lambda)$ is not forthcoming. In this article, we will present two novel connections between the minimization of $\Omega(\lambda)$ and information theory, one with coding theory (Cover & Thomas, 2006) and the other with modern model selection (Burnham & Anderson, 2002). As for the former, the minimization of $\Omega(\lambda)$ can be interpreted as a sort of compression problem of standard coding theory. As for the latter, the minimization of $\Omega(\lambda)$ resembles a kind of agnostic information theoretic model selection on the mapping of "forms" into "meanings".

Our optimization principles are called requirements in synergetic linguistics (Köhler, 2005). In synergetic linguistics the axioms of a theory of language include requirements and a special axiom, i.e. that language is a self-organizing and self-regulating system. Our justification of $\Omega$ will consist of two arguments. First, the justification of the ingredients of $\Omega(\lambda)$, namely the maximization of $I(S,R)$ (Section 3) and the minimization of $H(S)$ (Section 4). Second, a justification of their linear combination (Section 5) and some preliminary arguments on the self-organization that is required (Section 6).

3. THE PRINCIPLE OF MUTUAL INFORMATION MAXIMIZATION

Mutual information is a powerful measure of signal effectiveness (Bradbury & Vehrencamp, 2011). The next subsection reviews arguments for the maximization of mutual information between forms and meanings (Ferrer-i-Cancho, 2005a; Ferrer-i-Cancho & Díaz-Guilera, 2007).

3.1. *The maximization of I(S,R) promotes that forms become identifiers of meanings*

$I(S,R)$ is an information theoretic measure of the capacity of forms to convey meanings. Intuitively, the principle of mutual information maximization promotes somehow that forms become identifiers (IDs) of meanings (Ferrer-i-Cancho & Díaz-Guilera, 2007; Ferrer-i-Cancho, 2017; Ferrer-i-Cancho & Vitevitch, 2017). Put differently, the maximization of $I(S,R)$ is equivalent to the maximization of the expressivity of the system. When the number of meanings and forms is the same, it predicts one-to-one mappings between forms and meanings, which is intuitive to many researchers, but when there are more meanings than forms, optimal mappings may have disconnected meanings or consist of forms with multiple links (Ferrer-i-Cancho & Díaz-Guilera, 2007; Ferrer-i-Cancho, 2017; Ferrer-i-Cancho & Vitevitch, 2017).

3.2. *The maximization of I(S,R) sheds light on a vocabulary learning bias*

The principle of mutual information maximization sheds light on the origins of an intriguing vocabulary learning bias: when encountering a new word, children tend to assume that it means something totally different from the words that they already know (e.g., Markman & Wachtel, 1988; Merriman & Bowman, 1989; Clark, 1993). When a new word arrives, the child has to update his mapping of words into meanings. He/she has two options: a) Linking that word with an unlinked meaning and b) Linking the word with a linked meaning. Strategy a) is not only expected from the principle of contrast (Clark, 1987) and the more restrictive principle of mutual exclusivity (e.g., Markman & Wachtel, 1988; Merriman & Bowman, 1989) but also from mutual information maximization (Ferrer-i-Cancho, 2017; Ferrer-i-Cancho & Vitevitch, 2017).

3.3. *The maximization of I(S,R) promotes adherence to the principle of contrast safely*

The maximization of *I*(*S*,*R*) predicts the general principle of contrast (Clark, 1987) when the number of forms does not exceed the number of meanings (Ferrer-i-Cancho, 2017; Ferrer-i-Cancho & Vitevitch, 2017), which is apparently a realistic condition for human language. Notwithstanding, the principle of contrast does not warrant that forms convey enough information (unlinked forms do not violate the principle). In contrast, the maximization of *I*(*S*,*R*) imposes the further constraint over the principle of contrast that connections have to be distributed uniformly among forms particular models of Zipf's law, reducing the risk of unlinked forms (Ferrer-i-Cancho & Díaz-Guilera, 2007; Ferrer-i-Cancho, 2017; Ferrer-i-Cancho & Vitevitch, 2017). Thus, mutual information maximization is a powerful principle making realistic predictions about the mapping of forms into meanings.

3.4 *I*(*S*,*R*) *as a dual measure of cognitive cost*

The maximization of *I*(*S*,*R*) promotes that forms become identifiers (ID'S) of meanings. Expressing *I*(*S*,*R*) as a difference between unconditional entropies allows one to see that the maximization of *I*(*S*,*R*) benefits, both the speaker and the hearer (Ferrer-i-Cancho & Díaz-Guilera, 2007). On the one hand, one has

$$I(S,R) = H(S) - H(S \mid R), \tag{8}$$

which indicates that the maximization of *I*(*S*,*R*) promotes the reduction of *H*(*S*|*R*), the average entropy of forms given meanings, a measure of speakers cost as also suggested by Piantadosi (2014). If meanings have only one possible form (*H*(*S*|*R*) is minimum), producing the forms for the intended meanings is straightforward. On the other hand, one has

$$I(S,R) = H(R) - H(R \mid S), \tag{9}$$

which indicates that the maximization of *I*(*S*,*R*) promotes the reduction of *H*(*R*|*S*), the average entropy of meanings given forms, which is a measure of hearer's cost. If forms have only one possible meaning (*H*(*R*|*S*) is minimum), recovering the intended meanings is straightforward.

The maximization of *I*(*S*,*R*) promotes both the reduction of synonymy and the reduction of polysemy. The speaker's cost due to the synonymy of a meaning $r_j$ is captured by *H*(*S*|$r_j$), the entropy over the conditional probabilities of every form given the meaning (Ferrer-i-Cancho, 2005a, 2005c). The fact that a meaning has many synonyms does not imply a high cognitive cost because one of them could be used 99% of the times. Erasing any distributional information, that entropic cost can be roughly approximated by the number of synonyms, a metric that can be calculated rather simply. Indeed, standard information theory indicates that *H*(*S*|$r_j$) ≤ log $\omega_j$ where is $\omega_j$ is the number of forms for which *p*($s_i$,$r_j$)>0. In some models, one has *H*(*S*|$r_j$) = log $\omega_j$ (Ferrer-i-Cancho, 2005a, 2005c). Thus, $\omega_j$, the synonymy of a concept, defines an upper bound of the actual conditional entropy of a meaning or the actual conditional entropy in certain models. Psycholinguistic research on naming tasks has shown a disadvantage for synonymy: concepts with many forms are harder to name (Bates et al., 2003).

Symmetric arguments can be made for polysemy. The hearer's cost due to the polysemy of a form $s_i$ is captured by *H*(*R*|$s_i$), the entropy over the conditional probabilities of every meaning given the form (Ferrer-i-Cancho, 2005a, 2005c). The fact that a form has many meanings does not imply a high cognitive cost because one of them could be used 99% of the times. Erasing any distributional information, that entropic cost can be roughly approximated by the number of meanings, a metric that can be calculated rather simply. Indeed, standard information theory indicates that $H(R|s_i) \leq \log \mu_i$ where is $\mu_i$ is the number of meanings for which $p(s_i,r_j)>0$. In some models, one has $H(R|s_i) = \log \mu_i$ (Ferrer-i-Cancho, 2005a, 2005c). Thus, $\mu_i$, the polysemy of a form, defines an upper bound of the actual conditional entropy of a form or the actual conditional entropy in certain models.

Research on vocabulary learning shows that the potential polysemy of words increases over time in children (Casas, Català, Ferrer-i-Cancho, Hernández-Fernández & Baixeries, 2016) and second language speakers (Crossley, Salisbury, & McNamara, 2010), suggesting that words with lower polysemy are easier to learn. In contrast, an advantage for polysemous words, has been reported by psycholinguistic research on lexical decision tasks, where listeners are asked to indicate as quickly as they can if a sequence of letters is a real word, e.g., brick, or not, e.g., blick (Rubenstein, Garfield, & Millikan, 1970; Hino & Lupker, 1996; Kellas, Ferraro & Simpson, 1988; Millis & Button, 1989; Rubenstein, Lewis & Rubenstein, 1971). However, the advantage of more polysemous words could be a side-effect of the advantage for more frequent words in perception (Dahan, Magnuson, & Tanenhaus, 2001 and references therein) as more polysemous words tend to be more frequent (Zipf, 1945; Baayen, H. & Moscoso del Prado Martín, 2005; Ilgen & Karaoglan, 2007). Interestingly, polysemous words were responded to more quickly than words that are equivalent in frequency of occurrence but have only one meaning (Rubenstein et al., 1970). Such a non-trivial ambiguity advantage effect "seems to be counter-intuitive, as one might expect ambiguous words that have competing meanings to take longer to process" (Klepousniotou & Baum, 2007).

More precise psycholinguistics research has shown that indeed word ambiguity can be both advantageous and disadvantageous in lexical decision tasks depending on the kind of ambiguity: homonymy and polysemy (Rodd, Gaskell, & Marslen-Wilson, 2002). Homonymy is produced by different words that share the same orthographic and phonological form by chance, and their form has multiple unrelated meanings. Polysemy arises when a single word has various senses which are typically related (see Rodd et al., 2002, for further details and examples of each). So far, we have used the term polysemy as equivalent to ambiguity. It turns out that words with multiple senses are advantageous while words with multiple meanings are disadvantageous (Rodd et al., 2002). Therefore, the maximization of $I(S,R)$ captures advantages of polysemous words at the level of learning and a kind of ambiguity at the level of processing.

We suspect that a deeper insight into the effect of polysemy will be obtained when polysemy is measured by $H(R|s_i)$ instead of $\mu_i$ in psycholinguistic research. The same applies to $H(S|r_j) =$ versus $\omega_j$ for the effect of synonymy.

*3.5. General discussion*

The maximization of $I(S,R)$ corresponds to some extent to a mixture of minimization of production effort (minP) and the minimization of decoding effort (minD) in synergetic linguistics (Köhler, 1987; Köhler, 2005) at a semantic level. It is tempting to think that the principle of maximization of $I(S,R)$ could be replaced by a principle of minimization of $H(R|S)$ because $I(S,R)$ includes $H(R|S)$ (Eq. 9), a measure of decoding effort. $H(R|S)$, as a measure of the effort for the listener neglecting $I(S,R)$ (Corominas-Murtra, Fortuny, & Solé, 2011), is problematic: the minima of $H(R|S)$ are defined by mappings of forms into meanings where an arbitrary number of forms of non-zero probability are mapped into only one meaning each (Ferrer-i-Cancho & Díaz-Guilera, 2007; Prokopenko et al., 2010). Those minima can be easily deduced from the minima of $H(S|R)$ (Ferrer-i-Cancho & Díaz-Guilera, 2007) and swapping $S$ and $R$. The number of meanings that one is communicating about trough some form is irrelevant for the minimization of $H(R|S)$. Thus, either a one to one mapping of forms into meanings (a best case for listener effort) or a mapping where the only form that has non-zero probability is connected to a single meaning (a worst case for listener effort) are minima of $H(R|S)$. The danger of the minimization of $H(R|S)$ was neutralized imposing that meanings cannot be disconnected (Ferrer-i-Cancho & Solé, 2003). The problem is that this amendment compromises the parsimony of the theory. To conclude, $I(S,R)$ is not reducible to the conditional entropies $H(S|R)$ or $H(R|S)$.

Some researchers have regarded the minimization of $H(R|S)$ only as a benefit for the hearer (Corominas-Murtra et al., 2011). However, the minimization of $H(R|S)$ at constant $H(R)$ (Ferrer-i-Cancho & Solé, 2003) is equivalent to a maximization of $I(S,R)$ by virtue of Eqs. 8 and 9 and therefore thus implies pressure for minimizing $H(S|R)$, which can be regarded as reduction of the speaker's effort as explained above.

4. THE PRINCIPLE OF ENTROPY MINIMIZATION

The principle of form entropy minimization (Ferrer-i-Cancho & Solé, 2003; Ferrer-i-Cancho, 2005a; Ferrer-i-Cancho & Díaz-Guilera, 2007) is supported by theoretical and empirical arguments. Let us define $V_S$ as the number of forms that have non-zero probability ($V_S \leq V_S^{max}$).

Empirically, word entropy minimization is supported to some extent by Zipf's law (Eq. 2). Although vocabularies are far from minimum $H(S)$ ($H(S) = 0$) because they consist of more than one word, vocabularies are far from the maximum entropy because some words have a large frequency with regard to a large proportion of words occurring a few times. $\alpha \approx 0$ is obtained by the null hypothesis of choosing forms uniformly at random from a lexicon of finite $V_S$ (this prediction for $\alpha$ assumes $V_S > 1$; notice that $V_S = 1$ leads to $\alpha = \infty$). Thus, Zipf's law suggests some pressure for minimizing $H(S)$. The idea that the minimization of word entropy is not justified by the data (Piantadosi, 2014) is flawed.

The next subsections review a series of theoretical arguments ending with general discussion.

4.1 *H(S) as a measure of cognitive cost*

$H(S)$ is a measure of ease of production and perception at the level of word frequencies (Ferrer-i-Cancho & Díaz-Guilera, 2007). Psycholinguistic research indicates that the speed of accessing a phonological code is positively correlated with word frequency during speech production (Jescheniak & Levelt, 1994 and references therein). A similar effect is found in speech perception, where the speed with which a word is recognized is positively correlated with its frequency (Dahan et al., 2001, and references therein). $H(S)$ provides a measure of frequency-related costs in production and processing. First, it is minimum, i.e. $H(S)=0$, when only one form type can be produced ($V_S = 1$). It is maximum, i.e. $H(S) = \log V_S^{max}$, when all the form types are equally likely (Ferrer-i-Cancho & Díaz-Guilera, 2007). These two configurations constitute the most and the less favourable situation in terms of frequency related costs. When $V_S = 1$ the frequency of a form is maximized and the frequency related cost is minimized. When all forms are equally likely, they have their smallest probability and then their frequency related costs are the worst. Indeed, $H(S)$ can be expressed in terms of a Kullback-Leibler divergence of a uniform distribution ($p(s_i) = 1/V_S^{max}$) from the true form probabilities (Cover & Thomas, 2006, Section 2.7, p.32), i.e. [2]

$$H(S) = \log V_S^{\max} - D(p(s_i) \| 1/V_S^{\max}). \tag{10}$$

Thus $H(S)$ allows to measure of how far the vocabulary is from the worst case for frequency-related costs. From the principle that "children learn what they hear the most" (Konishi et al., 2014), we may infer that children tend to learn the most frequent words, and then, following the arguments above, conclude that $H(S)$ is also a measure of learning costs. All these arguments are fully consistent with a general statement, namely that "the frequency of a (linguistic) construct is a relevant factor determining its cognitive costs" (Fenk-Oczlon 2001, p. 435).

The statement that our models do not incorporate frequency effects or that the assumptions of the model are not justified on independent psychological grounds (Piantadosi, 2014) sweeps under the carpet psycholinguistic arguments in a series of articles (Ferrer-i-Cancho & Díaz-Guilera, 2007; Ferrer-i-Cancho & Solé, 2003) that have been update here.

### 4.2 $H(S)$ *as measure of the effective vocabulary size*

$H(S)$ is a measure of the effective vocabulary size that has to be minimized. Information theory defines the notion of the typical set, which is a subset of the vocabulary covering the overwhelming majority of the form occurrences, e.g., 99%. The size of this set is about $e^{H(S)}$ (Cover & Thomas, 2006). This size is indeed a measure of the effective vocabulary size (a repertoire could have $10^6$ types but if 1 of them is used 99% of the times, $10^6$ would be a very bad measure of the effective vocabulary size. The view of the entropy of forms as a powerful diversity measure is supported by the use of the entropy of species as an index of ecological diversity (Chao & Shen, 2003 and references therein). The minimization of the size of that typical set appears to be a powerful implementation of Zipf's principle of unification (Zipf, 1949) or the principle of minimization of memory effort (minM) in synergetic linguistics (Köhler, 1987; Köhler, 2005), which promote the minimization of the inventory size. Indeed, if the minimization of $H(S)$ is replaced by the minimization of the vocabulary size in optimization models of word frequencies, Zipf's law disappears (Ferrer-i-Cancho & Solé, 2003; Ferrer-i-Cancho, 2005a).

### 4.3 $H(S)$ *as a lower bound for compression*

The principle of compression (Chater & Vitányi, 2003; Ferrer-i-Cancho et al., 2013), which can be seen as a kind of energy minimization principle, could put indirect pressure to reduce $H(S)$. In a wide class of coding schemes, i.e. unique decipherable encoding, $H(S) \leq L(S)$, where $L(S)$ is the mean code length (Cover & Thomas, 2006). Therefore, reducing $H(S)$ increases the potential for compression. The latter implies that, a system may have to rearrange form probabilities to decrease the value of $H(S)$ so that $L(S)$ can be reduced further.

---

[2] Conditioning on $V_S$, it is obtained $H(S) = \log V_S - D(p(s_i) \| 1/V_S)$.

Outside unique decipherability, reducing $H(S)$ can help to reduce $L(S)$. Eventually, when $H(S) = 0$, only one form would have non-zero probability and then $L(S)$ could reach its absolute minimum. That means that the minimization of $H(S)$ is a way of facilitating compression neglecting to a large extent how forms are actually coded.

4.5 *The principle of contrast*

The minimization of $H(S)$ leads to a restricted version of the principle of contrast. That minimization also implies that $H(S|R)$ is being minimized indirectly, which leads to meanings that are linked with at most one form (Ferrer-i-Cancho & Díaz-Guilera, 2007). The point is that the minimization of $H(S)$ imposes the further constraint over the principle of contrast that connections have to concentrate on a single form.

4.6 *General discussion*

Eq. 7 suggests that the principle of entropy minimization can be regarded as maximization of the departure from the null hypothesis of a finite lexicon where forms are produced by choosing uniformly at random and thus $\alpha \approx 0$ (this is what McCowan, Hanser, & Doyle, 1999, call the zero-order level). Interestingly, this is the expected distribution of forms in the absence of any constraint except normalization (Janes, 1957).

The minimization of $H(S)$ is a sort of holistic requirement: it is beneficial learners and also for both hearer and speaker from a psycholinguistic perspective (but it does not include all the benefits for them). It integrates various requirements of synergetic linguistics (Köhler, 1987; Köhler, 2005) simultaneously: the minimization production effort (minP), the minimization of decoding effort (minD) and the minimization of memory effort (minM). Notice however that the minimization of $H(S)$ covers minP and minD only from the perspective of word frequency and minM from the perspective of lexicon size neglecting the cost of storing individual words. The memory cost depends on the coding scheme and $H(S)$ is only a lower bound to the mean code length (Cover & Thomas, 2006).

The view of $H(S)$ only as a benefit for the speaker (Corominas-Murtra et al., 2011) is outdated. It is inherited from the view of Zipf's view in the 1940s and must be updated with the discoveries of decades of psycholinguistic research and advances in information theoretic models of communication.

It has been argued that that $H(S|R)$ is a better choice than $H(S)$ as measure of speaker's difficulty since "this captures the uncertainty of the psychological system" (Piantadosi, 2014). $H(S)$ is also a measure of uncertainty of that system and interestingly it is not independent from $H(S|R)$: as $H(S) \geq H(S|R)$ (Appendix A), the minimization of $H(S)$ puts pressure for reducing $H(S|R)$. If $H(S)$ is minimum ($H(S)=0$) then $H(S|R)$ is also minimum ($H(S|R) = 0$) (Ferrer-i-Cancho & Díaz-Guilera, 2007). Put differently, the minimization of frequency related costs puts indirect pressure for the minimization of coding costs or synonymy related costs (recall our interpretation of $H(S|r_j)$ as a measure of synonymy related costs in Section 3.4). Finally, notice that the minimization of $H(S|R)$ is also an indirect consequence of the principle of maximization of $I(S,R)$ (recall Eq. 8).

## 5. THE STRUCTURE OF THE ENERGY FUNCTION

$\Omega(\lambda)$ is a cost function defined as linear combination of two principles: maximization of $I(S,R)$ and minimization of $H(S)$ (Eq. 7). These two principles are in conflict: minimum $H(S)$ implies minimum $I(S,R)$ as $I(S,R) \leq H(S)$ (Appendix A). Interestingly, the distribution of $H(S)$ is not symmetric but skewed to the right and show a gap between 0 and the minimum real value of $H(S)$ (Bentz et al., 2017), suggesting that pressure to maximize expressivity ($I(S,R)$) is limiting the pressure to minimize $H(S)$.

$\Omega(\lambda)$ has been used to explain the origins of Zipf's law (Eq. 2), the limits of the variation of α in languages (Ferrer-i-Cancho 2005b, Ferrer-i-Cancho, 2006) and a vocabulary learning bias (Ferrer-i-Cancho, 2017). However, one may still think that the design of $\Omega(\lambda)$ is arbitrary, lacking theoretical strength. For instance, it assumes that mutual information and form entropy must be combined linearly, but this raises the question of whether their combination is actually necessary or the question of whether a weighted sum of entropy and mutual information is unrealistic or naïve from an information theoretic perspective.

5.1. *The minimization of $\Omega(\lambda)$ makes powerful predictions in spite of its apparent arbitrariness*

In the domain where $H(S)$ is minimized, namely $0 \leq \lambda \leq 1/2$, the global minima satisfy the principle of contrast, namely that every two forms should have different meanings (Clark 1987):

- When $\lambda = 1/2$, the global minima of $\Omega(\lambda)$ of are those of $H(S|R)$, which implies that two forms cannot be linked with the same meaning (Ferrer-i-Cancho & Díaz-Guilera, 2007).
- When $0 \leq \lambda < ½$ the global minima are the minima of $H(S)$, but the fact that $H(S) \geq H(S|R)$ (Appendix A) implies that the global minima of $H(S)$ are subset of the global minima of of $H(S|R)$ and thus configurations following the principle of contrast are obtained again (Ferrer-i-Cancho, 2017; Ferrer-i-Cancho & Vitevitch, 2017).

Notice that the global minima of $\Omega(\lambda)$ are unrealistic for not warranting successful communication ($I(S,R)>0$) and not yielding Zipf's law (when $\lambda<1/2$, only one form is predicted and thus α=∞) but might explain the attraction towards the principle of contrast to some extent.

The minimization of $\Omega(\lambda)$ also allows one to predict the tendency of children to attach new words to unlinked meanings. Interestingly, the prediction does not depend on the choice of $\lambda$, provided that $0 \leq \lambda \leq 1$ (Ferrer-i-Cancho, 2017). Finally, the minimization of $\Omega(\lambda)$ is able to reproduce, besides Zipf's law for word frequencies, Zipf's meaning-frequency law qualitatively (in particular, Eq. 1 with δ = 1; Ferrer-i-Cancho, 2016a). If the design $\Omega(\lambda)$ was linguistically or psychologically superficial or if its design was arbitrary, why is it able to predict various patterns of language? The answer may be in two $\Omega(\lambda)$ independent links with information theory that are presented in the next subsections.

## 5.2 *The minimization of Ω(λ) as compression*

In our setup, forms from *S* code for meanings in *R*. Then *R* is the source set of symbols and *S* is the target set of symbols of the mapping of meanings into forms. In standard coding theory, the problem of compression consists of minimizing *L(S)*, the mean length of the forms in *S*, given the probability of each meaning under some coding scheme, namely some assumptions on the mapping above (Cover & Thomas, 2006). An elementary assumption is that no meaning can be left uncoded (the function that translates a meaning into a form must be total). On top of it, a general assumption is non-singular coding, namely, that every meaning of *R* maps into a different string in *S*, namely the mapping above must be injective (Cover & Thomas 2006, pp. 105). Therefore, the standard problem of compression can be recast as the problem of minimizing

$$\Psi(\lambda) = J(S,R) + L(S), \tag{11}$$

where *J(S,R)* = 0 if the coding satisfies the conditions of the coding scheme and *J(S,R)* = ∞ otherwise. Suppose that the scheme are non-singular codes. Then, *J(S,R)* is a cost function whose minima are non-singular mappings.

We can generalize the standard problem of compression replacing *J(S,R)* by a cost function whose minima are still non-singular mappings but that exhibits a greater capacity of variation. A simple alternative is *H(R|S)* because its minima imply that every form is mapped to at most one meaning (Prokopenko et al., 2010; Ferrer-i-Cancho & Díaz-Guilera, 2007). While *J(S,R)* can only take two values (0 or ∞), *H(R|S)* has the capacity to vary between 0 and log $V_R^{max}$ (Cover & Thomas, 2006). This allows one to transform the definition of Ψ(λ) in Eq. 11 into one where the two constraints, i.e. the minimization of *H(R|S)* and the minimization of *L(S)*, are combined linearly as

$$\Psi(\lambda) = \lambda H(R|S) + (1-\lambda)L(S). \tag{12}$$

If the cost function is *H(R|S)* one still has to impose the elementary assumption i.e., no meaning can be left uncoded. An objective function that gives the elementary condition and a particular case of non-singular coding as maxima is *I(S, R)* when $V_S^{\max} \geq V_R^{\max}$ (Ferrer-i-Cancho, 2017; Ferrer-i-Cancho & Vitevitch, 2017). In particular, the maxima of *I(S, R)* in models where form probability is a power-function of its number of connections are defined by a couple of conditions (Ferrer-i-Cancho, 2017; Ferrer-i-Cancho & Vitevitch, 2017):

1. All forms are equally likely (they all have the same number of connections).
2. Every meaning is connected to at most one form (principle of contrast).

when $V_S^{max} \leq V_R^{max}$. By symmetry, the configurations that maximize *I(S,R)* when $V_S^{max} \geq V_R^{max}$ are defined by the conditions:

1. All meanings are equally likely (they all have the same number of connections).
2. Every form is connected to at most one meaning.

Replacing the minimization of *H(R|S)* by the maximization of *I(S, R)* in Eq. 12, it is possible redefine Ψ(λ) as

$$\Psi(\lambda) = -\lambda I(S,R) + (1-\lambda)L(S). \tag{13}$$

The scheme of uniquely decipherable coding is a particular case of non-singular coding. Assuming optimal non-singular coding, $\Psi(\lambda)$ becomes $\Omega(\lambda)$ (Eq. 7) because $L(S) \approx H(S)$ under optimal uniquely decipherable coding (Cover & Thomas, 2006). Similarly, if $L(S)$ is replaced by $H(S)$ in Eq. 12, one obtains

$$\Omega(\lambda) = \lambda H(R|S) + (1-\lambda)H(S), \tag{14}$$

namely the cost function of a variant of the optimization model (Ferrer-i-Cancho & Solé, 2003) that is equivalent to Eq. 7 when $H(R)$ is constant thanks to Eq. 9.

Therefore we can reinterpret the minimization of $\Omega(\lambda)$ to the light standard coding theory as a sort of compression that is obtained from a relaxation of non-singular coding and the assumption of optimal uniquely decipherable coding. These results are reminiscent of the recent suggestion that Zipf's law for word frequencies could originate from simultaneous pressure for optimal non-singular coding and optimal uniquely decipherable coding (Ferrer-i-Cancho, 2016b). The minimization of $H(S)$ can also be interpreted as the maximization of the potential for compression under uniquely decipherable encoding, since $L(S) \geq H(S)$ for this coding scheme (Cover & Thomas, 2006, p.111). The minimization of $H(S)$ may also help to reduce $L(S)$ in other coding schemes. Finally, notice that the optimization setup based on Eq. 7 also deviates from standard information theory because meaning probabilities are free whereas they are given (fixed) in the standard problem of compression. With this respect, the variant of the model based on Eq. 14 (Ferrer-i-Cancho & Solé 2003) is closer to standard information theory for assuming that meaning probabilities are constant (in particular, all meanings are equally likely).

So far we have justified the combination of $I(S,R)$ and $H(S)$ in $\Omega(\lambda)$ that was chosen to be linear for the sake of simplicity. Next section speculates on a possible rationale for such an additive combination.

5. 3 *The minimization of $\Omega(\lambda)$ as agnostic model selection*

After a quick introduction to information theoretic model selection (Burham & Anderson, 2002), the relationship between the minimization of $\Omega(\lambda)$ will be reviewed. Imagine that one wishes to model a sample of numbers through a distribution function *g* being *f* the true distribution. Information theoretic model selection is concerned about minimizing $D(f \| g)$, the Kullback-Leibler divergence of *f* from *g* that is also expressed as a divergence from *g* to *f*. This wording reflects the view of *f* as the posterior distribution and *g* as the prior distribution. $D(f \| g)$ "denotes the information lost when *g* is used to approximate *f*" (Burham & Anderson, 2002; p. 51). The best approximating model *g* is the model with smallest $D(f \| g)$. One of the major components of that theoretical framework are rather simple measures whose minimization is equivalent (under certain general conditions) to the minimization of $D(f \| g)$. One of the most popular measures is AIC, Akaike's information criterion, defined as (Burham & Anderson, 2002),

$$AIC = -2\log L + 2K, \tag{15}$$

where *L* is the maximum likelihood given the data and a model *g* and *K* is the number of parameters of model *g*. The best model is the model minimizing AIC. Interestingly, AIC and similar metrics, combine the minimization of -log *L*, a measure of the quality of the fit of a model (the lower its value, the higher the quality), with a minimization of *K*, which favours parsimony. Thus, the goal of information theoretic model selection turns out to be the quest for a model *g* that minimizes a linear combination of the quality of the fit of the model and a penalty for the number parameters of *g*. It is important to notice that parsimony is not an assumption but a by-product of the minimization of *D*(*f* || *g*) (Burham & Anderson, 2002; p. 63). Put differently, Occam's razor is not an assumption but a consequence for information theoretic model selection.

First we will examine the relationship between the minimization of $\Omega(\lambda)$ and the minimization of *D*(*f* || *g*) and then we will examine the relationship between the minimization of $\Omega(\lambda)$ and the minimization of AIC. As for the former, notice that *I*(*S*,*R*) is a Kullback-Leibler divergence from the joint probability of forms and meanings under independence ($p(s_i)p(r_j)$) to the actual joint probability ($p(s_i,r_j)$), i.e.

$$I(S,R) = D(p(s_i,r_j) \| p(s_i)p(r_j)). \tag{16}$$

This equivalence is easy to see by means of the definition of *I*(*S*,*R*) in Eq. 6. In Eq. 16, $p(s_i,r_j)$ can be interpreted as *f*, the true distribution and ($p(s_i)p(r_j)$) can be interpreted as model *g* which is a function of *f* thanks to Eqs. 3 and 4. Thus, the minimization of $\Omega(\lambda)$ when $\lambda > 0$ implies a maximization of *I*(*S*,*R*), the Kullback-Leibler divergence from random independent form-meaning associations to their actual joint probability. The maximization of *I*(*S*,*R*) resembles an agnostic model selection:

- No bet is made on the best candidate model *g*. Indeed, the *g* proposed is the worst. *g* = $p(s_i)p(r_j)$ is the model to avoid.
- There is no specific "true" distribution *f* for reference (the "true" distribution *f* = $p(s_i,r_j)$ is allowed to vary) but the "true" distribution must be as far as possible from a random model *g*.

Instead of minimizing the divergence of $p(s_i,r_j)$ from a reference model (the "true" distribution) as in a typical model selection setup, mutual information maximization maximizes the divergence of the null hypothesis that forms and meanings are independent (the worst model) from $p(s_i,r_j)$, the "true" distribution. Mutual information does not indicate how good a certain model is but what reality (the true distribution) should scape from. Mutual information maximization is agnostic about the true *f* = $p(s_i,r_j)$ while the "candidate model" *g* = $p(s_i)p(r_j)$ is fixed. For these two reasons, the agnostic model selection imposed (Eq. 16) may need to be complemented by the minimization of *H*(*S*) in $\Omega(\lambda)$ to obtain a parsimonious model, a justification for the minimization of *H*(*S*) that has not been considered in the reductionist review of justifications of the minimization of *H*(*S*) in Section 4. Thus, the minimization of $\Omega(\lambda)$ when $0<\lambda<1$ resembles a combination of agnostic model selection with the further constraint of minimizing *H*(*S*) (Section 4).

As for the relationship between the minimization of AIC and the minimization of Ω(λ), notice that the structural similarity between AIC and Ω(λ) is striking: both AIC and Ω(λ) follow the same pattern: a measure of quality (*I*(*S*,*R*) or -log *L*) is combined linearly with a penalty to favour parsimony (*H*(*S*) or *K*). While information theoretic model selection favours the quality of the fit of the model by maximizing the log-likelihood (Eq. 15), information theoretic models of communication favour the quality of the communication system by maximizing the mutual information. Indeed log-likelihood and Kullback-Leibler divergence (e.g., mutual information) bear a close relationship from an information theoretic perspective. $D(h \| g)$, the Kullback-Leibler divergence from a theoretical distribution *g* (*g* is a theoretical model) to an empirical distribution *h* satisfies

$$D(h \| g) = -\log \overline{L}, \qquad (17)$$

where $\overline{L} = L^{1/n}$ is the average likelihood with *n* as the sample size (Shlens, 2007). For instance, *g* can be a probability distribution for a certain random variable (e.g., a zeta distribution) and *h* can be the relative frequency of each value of a random variable in a sample (Shlens, 2007).

On the one hand, the minimization of AIC derives from the minimization of a Kullback-Leibler divergence from a candidate model *g* to true distribution *f* (Burnham & Anderson, 2002) that implies the minimization of a Kullback-Leibler divergence from a candidate model *g* to an empirical distribution *h* (Eq. 17). On the other hand, the minimization of Ω(λ) implies the maximization of a Kullback-Leibler divergence (*I*(*S*,*R*)). Those differences does not imply that that AIC and Ω(λ) are radically different: the maximization of mutual information can be seen as the minimization of the likelihood of an "empirical" distribution *h* with respect to a theoretical distribution *g* (Eq. 17), where *h* would be the joint probability of forms and meanings (*p*(*s*$_i$,*r*$_j$)) and *g* would be the joint probability of those mapping under statistical independence (*p*(*s*$_i$)*p*(*r*$_j$)). Put differently, the maximization of mutual information can be seen as the maximization of the unlikelihood according to an arbitrary mapping of forms into meanings. While information theoretic model selection favours parsimony by minimizing *K*, information theoretic models of communication favour parsimony by minimizing *H*(*S*), which is a measure of the effective vocabulary size, the size of the typical vocabulary, or put differently, the effective vocabulary size (Section 4.2). Another interpretation for the minimization of *H*(*S*) as a pressure for parsimony is that the minimization of *H*(*S*) increases the potential for compression (Section 4.3). The minimization of *H*(*S*) is a way of favouring parsimony that is to a large extent neutral about the coding scheme actually used to map meanings into forms.

In spite of the many parallels between AIC and Ω(λ), the matching is not perfect for various reasons, some discussed above. Another reason is that AIC gives equal weight to both –log *L* and *K*, whereas Ω(λ) only gives equal weight to *H*(*S*) and –*I*(*S*,*R*) when λ = 1/2. Interestingly, there are variants of AIC that do not give equal weight to each of the ingredients (Burharnm & Anderson, 2002). A useful variant is the AIC corrected for samples of small size *n* (Burharnm & Anderson, 2002), that can be defined as

$$AIC_c = -2\log L + 2K \frac{n}{n - K - 1}, \qquad (18)$$

where *n* is the sample size. On the one hand, the minimization of AIC$_c$ is equivalent to the minimization of (Appendix B)

$$D(h \| g) + a_{AIC} K, \qquad (19)$$

where *h* is an empirical distribution, *g* is a theoretical distribution (a candidate model) and $a_{AIC}$ is a parameter ranging from 0 to ∞. On the other hand the minimization of Ω($\lambda$) is equivalent to the minimization of (Appendix B)

$$-D(p(s_i, r_j) \| p(s_i) p(r_j)) + a_\Omega H(S), \qquad (20)$$

where $a_\Omega$ is a parameter ranging from 0 to ∞.

Comparing Eqs. 19 and 20, the similarity between the minimization of AIC and that of Ω($\lambda$) becomes clearer. Fist, the two factors that control the weight of parsimony, i.e. $a_{AIC}$ and $a_\Omega$ respectively, range from 0 to ∞ (see Appendix B for further details). Second, $a_{AIC}$ depends on *n*, the sample size (Appendix B), while $a_\Omega$ depends on $\lambda$. Interestingly, the critical value of $\lambda$ where Zipf's law emerges may depend on the size of the joint probability matrix (Ferrer-i-Cancho, 2005a), thus introducing a dependency on $V_S$ and $V_R$. Therefore, we conclude that the minimization of AIC and Ω($\lambda$) exhibits a strikingly similar mathematical structure (Eqs. 19 and 20). A key formal difference is the sign of *D*, which is positive in Eq. 19 and negative in Eq. 20.

Parallel conclusions can be reached for other information theoretic criteria, e.g., the Bayesian Information Criterion (BIC) defined as (Burham & Anderson, 2002, p. 271)

$$BIC = -2 \log L + \log(n) K \qquad (21)$$

with *n* as the sample size. BIC bears some specific similarities with Ω($\lambda$). BIC introduces a penalty for parsimony that depends logarithmically on the size of the sample (Eq. 21) while Ω($\lambda$) introduces a similar penalty through the minimization of *H(S)* that is bounded above by log $V_S^{max}$.

The minimization of Ω($\lambda$) using a Monte Carlo procedure at zero temperature leads to a distribution consistent with Zipf's law for $\lambda^*$, a critical value of $\lambda$ close to 1/2 (Ferrer-i-Cancho & Solé, 2003; Ferrer-i-Cancho, 2005a) suggesting, to the light of the current article, that form frequencies result from a critical balance between parsimony and maximizing a "distance" to independence. It is not surprising that languages that differ tremendously from a typological perspective exhibit Zipf's law (Zipf, 1949). The minimization of Ω($\lambda$) resembles a theory agnostic model selection that favours parsimony abstracting away from the coding scheme actually used.

5.3 *General discussion*

We have stablished a connection between the minimization of Ω($\lambda$) and the problem of compression and speculated on another connection with model selection. We have argued that is reasonable to combine –*I(S,R)* and *H(S)* linearly in a direct fashion. If the minimization of *H(S)* is taken literally as pressure for reducing the effective size of the inventoire as suggested in Section 4, one feels tempted to replace *H(S)* by $e^{H(S)}$ in the definition of Ω($\lambda$). However, adjustments to Ω(λ) might be necessary. Notice that the scales of *I*(*S*,*R*) and *H*(*S*) are the same in the original Ω(λ) because *I*(*S*,*R*)≤*H*(*S*). This new variant of Ω(λ) should be the subject of future research.

We have shown that the Kullback-Leibler divergence underlies the minimization of *H*(*S*) and also the maximization of *I*(*S*,*R*). While the former maximizes a "distance" to equiprobability (Section 4.1), the latter maximizes a "distance" to a random mapping of forms into meanings (Section 5.3). Kullback-Leibler divergence has also been used to investigate the origins of Zipf's law for word frequencies (Corominas-Murtra et al., 2011), i.e. through a set of constraints, including the minimization of the divergence from the next word probability distribution to the current distribution (in this setup, the prior is the next word probability distribution, contrary to one's expectation in a Bayesian framework). In contrast, our framework is based on Kullback-Leibler divergence maximization from disorder.

6. SELF-ORGANIZATION

We have unveiled a striking similarity between optimization of communication (through Ω(λ)) and information theoretic model selection. In that framework, AIC is defined as a linear combination of terms (Eqs. 15 or 18), where the weight of each is not arbitrary but derived theoretically: the relative weight of log-likelihood with respect to *K* in AIC is important (Burnharm & Anderson, 2002, p. 64). The same applies to other metrics. The minimization of Ω(λ) is based on a linear combination of -*I*(*S*,*R*) and *H*(*S*) where the relative weight of each is determined by λ. Thus, an important difference between AIC-like metrics and Ω(λ) is that Ω(λ) leaves the issue of the right weight of the terms open.

6.1. *The hidden complexity of Simon's model*

In spite of the many predictions that Ω(λ) is able to make (Zipf's law, principle of contrast, a vocabulary learning bias), it is possible to remain skeptical about the power of the minimization of Ω(λ) because it yields Zipf's law only for a critical value of λ below ½ (Piantadosi, 2014) and, according to many researchers, Zipf's law for word frequencies can be reproduced simply (e.g., Miller & Chomsky, 1963). Until recently, random typing has been the typical simple mechanism invoked, but detailed statistical investigation has revealed that the model does not fit the data as promised (Ferrer-i-Cancho & Elvevåg, 2010; Ferrer-i-Cancho & McCowan, 2012). An alternative simple explanation is Simon's model, which produces a sequence of forms by selecting a form that has already been produced with probability α and produces a new form with probability 1-α. The old form is chosen uniformly over all the tokens of the sequence. This means that the probability of choosing a certain form is proportional to its current frequency. Interestingly, Simon's model reproduces Zipf's law (Eq. 2) with an exponent α. Occam's razor could favour Simon's model over our more sophisticated information theoretic model based on the optimization of Ω(λ) but one has to make sure that the definition of Simon's model does not require a careful choice of assumptions.

Let us consider that a generalization of Simon's model where the probability of choosing a certain form is proportional to $f^\phi$ where ϕ is a parameter (Chung, Handjani, & Jungreis, 2002). The outcome of the dynamics of the model depends on ϕ:

- If ϕ = 1 one has the original Simon model and Zipf's law is obtained as before.
- If ϕ > 1 a single form dominates.
- If ϕ < 1 (with parameter α < 1), an exponential distribution of form frequencies is obtained.

Thus, Simon's model hides a delicate assumption about the dependency between the probability of choosing a form and its frequency. The point is how ϕ could be tuned naturally. Self-organization could be the answer.

The concept of self-organization is crucial for complex thinking (Morin, 1990). Indeed, *"a central axiom of synergetic linguistics is…that language is a self-organizing and self-regulating system"* (Köhler, 2005). The problem of the delicate assumptions of the original Simon's model can easily fixed viewing language as a complex self-organizing system (Oudeyer, 2006; Köhler, 2005; Steels, 2000; Köhler, 1987). A self-organizing communication system based on the dynamical principles of the generalized Simon model might easily choose ϕ ≤ 1 if it was penalized by the fact that when a single form dominates, communication is not possible because $H(S) = 0$ and thus $I(S,R) = 0$. ϕ might converge to 1 adding the further constraint that $H(S)$ must be reduced. The patches that Simon's model needs are the definition of the model of Zipf's law based on the minimization of $\Omega(\lambda)$. Simon's model is not as simple as commonly believed (Rapoport, 1982). However, this conclusion depends on the assumption that the probability of choosing a certain form can only be proportional to $f^\phi$, independently of time or text length. The actual functional dependency should be investigated.

6.2 *A communication system minimizing $\Omega(\lambda)$ could self-organize*

Let us forget about Simon's model. A communication system self-organizing through the minimization of $\Omega(\lambda)$ could find the right value of $\lambda$ by its own and naturally. When $\lambda=0$ communication is impossible (Eq. 7 indicates that $I(S,R)$ is irrelevant and minimum $H(S)$ implies $H(S) = I(S,R) = 0$; Appendix A). When $\lambda=1$ communication is perfect ($I(S,R)$ is maximized freely) but at a maximum cost (Eq. 7 indicates that $H(S)$ is irrelevant). The minimization of $H(S)$ tends to dominate for $\lambda < ½$ while the maximization of $I(S,R)$ tends to dominate for $\lambda > ½$ (Ferrer-i-Cancho & Solé, 2003; Ferrer-i-Cancho, 2005a). The proper value of $\lambda$ could be easy to determine through feedback from the current choice of $\lambda$. Recall that the minimization of $H(S)$ and the maximization of $I(S,R)$ are in conflict because $I(S,R) \leq H(S)$ (Appendix A). Although the minimization of $H(S)$ would eventually lead to $H(S)=0$ that would reduce expressivity ($I(S, R)$) to zero. The gap between 0 and the minimum real value of $H(S)$ (Bentz et al., 2017) suggests that languages could be self-organizing to preserve expressivity.

6.3 *Self-organizing at the vicinities of a critical point looks easy for natural systems*

One may argue self-organizing at the vicinities of a critical point is still a hard problem (Piantadosi, 2014). However, this difficulty might be just a premature conclusion of our limited understanding of the dynamics and constraints of the real problem, not an intrinsic property of critical points. Indeed, the brain shows spatiotemporal patterns of criticality in the resting state, indicating that this criticality can be reached and mantained "effortlessly" (Haimovici, Tagliazucchi, Balenzuela, & Chialvo, 2013). Recently, Hidalgo, Grilli, Muñoz, Banavar and Maritan (2015) have shown the evolutionary advantage of poising a system in the vicinities of a critical point from a theoretical perspective. A lesson from information theoretic models selection is that the relative weight of quality of fit (likelihood) with respect to parsimony (the number of parameters) cannot be any: it depends on properties of the samples (e.g., sample size) or assumptions about the models being evaluated (Wagenmakers & Farrell, 2004). An incorrect weighting can lead to wrong statistical inferences (Burnharm & Anderson, 2002, p. 64). After all, the diversity of adequate weightings provides support for the suitability of a model of communication requiring a precise tuning of $\lambda$ to perform an optimization that resembles a kind of compression or agnostic model selection. In sum, flexibility can be a strength.

7. DISCUSSION

From Sections 3 and 4 we can draw some conclusions: the principle of minimization of $H(S)$ is (at least) a principle of the minimization of frequency related costs while the principle of the maximization of $I(S,R)$ is (at least) a principle of the minimization of polysemy/synonymy related costs. Researchers have been worried about splitting costs into speaker's and hearer's (Zipf, 1949; Ferrer-i-Cancho & Solé, 2003; Corominas-Murtra et al., 2011) but careful analysis (here and also Ferrer-i-Cancho & Díaz-Guilera, 2007) shows that the two fundamental information theoretic principles above concern both sides. Perhaps the key is not to whom a cost belongs but the kind of cost. Thus the minimization of $\Omega(\lambda)$ can be seen as a reduction of frequency and polysemy/synonymy related costs. Interestingly, the minimization of $\Omega(\lambda)$ is linked to the problem of compression of standard coding theory and also resembles a kind of agnostic model selection that abstracts away from the coding scheme. That might explain the capacity of $\Omega(\lambda)$ minimization to explain various patterns of language, including Zipf's law for word frequencies. However, we have just suggested a possible relationship between the minimization of $\Omega(\lambda)$ and model selection. Further mathematical work is needed to establish a stronger connection with model selection.

Our considerations about Zipf's law for word frequencies and its origins can be framed into the *"general debate between explanations that consider widespread scaling laws as coincidences due to multifactorious origins, versus explanations that consider scaling laws as expressions of general principles of cognitive function at both neural and behavioural levels of analysis. The same basic debate between domain-specific versus domain-general explanations of scaling laws has been unfolding throughout the sciences for decades, suggesting a deep issue at stake"* (Kello, 2013). We believe that the proponents of domain-specific explanations, who have been seduced by the local simplicity of specific models, have not worried sufficiently about the implications of their oversimplification and reductionism for the construction of compact general theories beyond the target scaling law. Not to mention that those simplistic models in some cases do not meet the requirement of providing a sufficiently good fit to the real data (Ferrer-i-Cancho & Elvevåg, 2010; Ferrer-i-Cancho & McCowan, 2012; Ferrer-i-Cancho, Hernández-Fernández, Baixeries, Dębowski, & Macutek, 2014). Some specific arguments for Zipf's law for word frequencies will be presented next.

It might be true that there are many ways of reproducing power-laws such as Zipf's law for word frequencies, but there are not so many models that allow one to explain the origins of the law, the principle of contrast, a vocabulary learning bias,…in one shot. The idea that the family of models reviewed here does not make predictions beyond Zipf's law or that it has not been tested with new independent data (Piantadosi, 2014) is flawed.

A researcher who wishes to provide an explanation of Zipf's for word frequencies that is constrained at least by the other patterns reviewed in Section 1 will have to go to the market of science and consider at least two kinds of products. Firstly, heavy packages like one that includes the following items:

- A model of Zipf's law that does not belong to the family of models reviewed here.
- The principle of contrast (to explain synonymy avoidance) or a model that explains it.
- A model to explain the meaning-frequency law, at least qualitatively.
- Mutual information maximization, which is needed to warrant successful communication (Ferrer-i-Cancho, 2017; Bradbury & Vehrencamp, 2011).

The model of Zipf's law might be sums of random variables from heavy tailed distributions (Willinger, Alderson, Doyle, & Li, 2003). There are even simpler ways of producing a power law distributions such as inverting uniform random numbers but they give concrete exponents ($\alpha = 1$ in the inversion example) that are too restrictive for the values of the power-law exponent $\alpha$ that are found in real word frequencies (Ferrer-i-Cancho & Servedio, 2005; Baixeries et al., 2013).

If the model of Zipf's law is Simon's model with $\delta = 1$, the scientist may have to buy some extra products to obtain $\delta = 1$, such as

- Entropy minimization (mutual information maximization is already included in the original package; recall the arguments in Section 6 for Simon's model).
- Self-organization (Section 6).

Alternatively, the scientist may also have to consider a lighter package containing

- A model of Zipf's law from the family of models reviewed here.
- Self-organization (Section 6).

Notice that in the lighter package, the model of Zipf's law can be an item that is heavier than the model of Zipf's law in the heavy package. In the lighter package, the weight of a model of Zipf's law from the family of models reviewed here, which is due to a combination of entropy minimization (Section 3) and mutual information maximization (Section 4), is compensated by its modelling versatility (predictive power), allowing one to reproduce various language patterns.

The deep scientist will have noticed that the packages of models, specially the heavy ones, may not constitute a scientific theory. Scientific knowledge is more than an ensemble of models. As Bunge puts it, *"Scientific knowledge is systematic: a science is not an aggregation of disconnected information, but a system of ideas that are logically connected among themselves. Any system of ideas that is characterized by a certain set of fundamental (but refutable) peculiar hypotheses that try to fit a class of facts is a theory"* (Bunge, 2013, pp. 32-33). Research on disconnected little models is good to increase one's publications list and maximize the chances of success in a system where quantity is mistaken for quality and interest in the development of general theories is seen as naive (for lacking domain specific knowledge) or too ambitious. However, these sociological pressures are a serious hindrance for the collective project of increasing scientific knowledge: the risk of the formation of fat, superficial and disorganized pseudoscientific theories at a larger scale is higher.

In the coming years, researchers will have to choose between

- Simple but unrealistic explanations of Zipf's law for word frequencies that lead to fat hypothesis when integrated within a tentative general theory of communication (with language as particular case), cognition or the functioning of natural systems, or
- More sophisticated models of Zipf's law that can handle a wide range of abstract phenomena, not necessarily cognitive or human, elegantly and with a little extra cost.

A new avenue for research is offered by a generalization the model in Ferrer-i-Cancho (2005) that allows one to reproduce the meaning-frequency law with the right exponent, namely Eq. 1 with $\delta$=0.5 (Ferrer-i-Cancho & Vitevitch, 2017). Thes new generation promises to increase the predictive power of the family of optimization models reviewed in this article.

Beyond Zipf's law and language laws, researchers will have to choose between

- Independent models for human language, animal communication and genomes that lead to a fat global theory when merged or
- Theoretical approaches working at a level of abstraction where it is possible to develop a compact but still deep knowledge about life.

APPENDIX A

The fact that $I(S, R) \geq 0$ (Cover & Thomas, 2006) implies that $H(S) \geq H(S|R)$ (thanks to Eq. 8) and also that $H(R) \geq H(R|S)$ (thanks to Eq. 9). As $H(S|R), H(R|S) \geq 0$ we also have $I(S, R) \leq H(S), H(R)$ (via Eqs. 8 and 9). Then it is easy to see that $H(S) = 0$ or $H(R) = 0$ imply $I(S, R)=0$.

APPENDIX B

Applying $\overline{L} = L^{1/n}$ (Shlens, 2007) to Eq. 16, the AIC corrected for finite samples becomes

$$AIC_c = -2n \log \overline{L} + 2K \frac{n}{n - K - 1}. \tag{B1}$$

Thanks to Eqs. 15 and B1, the minimization of AIC$_c$ is equivalent to the minimization of

$$AIC_c /(2n) = D(h \| g) + a_{AIC} K \tag{B2}$$

with

$$a_{AIC} = \frac{1}{n - K - 1}. \tag{B3}$$

Requiring $n \geq K + 1$ it is obtained that $a_{AIC}$ ranges from 0 to ∞. This constraint is justified by the need that the sample is large enough with regard to number of parameters. In non-linear regression, it is customary to require $n \geq K + 1$, being $K$ the number of explicit parameters of the model and the +1 due to the assumption of noise with 0 mean and constant variance (Ritz & Strebig, 2008, p. 1).

Combining Eqs. 7 and 14, it follows that the minimization of $\Omega(\lambda)$ is equivalent to the minimization of

$$\Omega(\lambda)/\lambda = -D(p(s_i, r_j) \| p(s_i)p(r_j)) + a_\Omega H(S) \tag{B4}$$

with

$$a_\Omega = \frac{1}{\lambda} - 1. \tag{B5}$$

Recalling that $0 \leq \lambda \leq 1$, it is is easy to see that $a_\Omega$ ranges from 0 to $\infty$.

ACKNOWLEDGEMENTS

We thank C. Bentz, C. Kello, A. Hernandez-Fernández, E. Santacreu-Vasut and M. Vitevitch for their generous advice. We are also grateful to G. Bel-Enguix, M. Christiansen, J. M. Fontana, A. Hernández-Fernández, L. McNally, F. Moscoso del Prado Martín, S. Piantadosi and E. Santacreu-Vasut for helpful discussions. Core ideas of this article were presented at the investigative workshop, "Analyzing Animal Vocal Communication Sequences" that took place on October 21-23 2013 in Knoxville, Tennessee, sponsored by the National Institute for Mathematical and Biological Synthesis (NIMBioS). This work was supported by the grant 2014SGR 890 (MACDA) from AGAUR (Generalitat de Catalunya) and also the APCOM project (TIN2014-57226-P) from MINECO.